\documentclass[11pt,twoside]{article}

\usepackage{asp2014}

\aspSuppressVolSlug
\resetcounters

\begin{document}

\title{New Python-based Architecture for the Keck Observatory Archive}

\author{R.~Moseley,$^1$ G.~Bruce~Berriman,$^1$ Christopher~R.~Gelino,$^1$ 
 John~C.~Good,$^1$ and Toba~Oluyide$^1$}
\affil{$^1$Caltech/IPAC-NExScI,~Pasadena,~CA~91125,~USA; \\
\email{moseley@ipac.caltech.edu}}

\paperauthor{R.~Moseley}{moseley@ipac.caltech.edu}{0000-0002-0759-0475}{Caltech}{IPAC/NExScI}{Pasadena}{CA}{91125}{USA}
\paperauthor{G.~Bruce~Berriman}{gbb@ipac.caltech.edu}{0000-0001-8388-534X}{Caltech}{IPAC/NExScI}{Pasadena}{CA}{91125}{USA}
\paperauthor{C.~R.~Gelino}{cgelino@ipac.caltech.edu}{0000-0001-5072-4574}{Caltech}{IPAC/NExScI}{Pasadena}{CA}{91125}{USA}
\paperauthor{J.~C.~Good}{jcg@ipac.caltech.edu}{0009-0003-3906-719X}{Caltech}{IPAC/NExScI}{Pasadena}{CA}{91125}{USA}
\paperauthor{T.~Oluyide}{toluyide@ipac.caltech.edu}{}{Caltech}{IPAC/NExScI}{Pasadena}{CA}{91125}{USA}

\begin{abstract}
We describe the development of the Keck Observatory Archive (KOA) Data Discovery Service, a web-based dashboard that returns metadata for wide-area queries of the entire archive in seconds. Currently in beta, this dashboard will support exploration, visualization, and data access across multiple instruments. This
effort is underpinned by an open-source, VO-compliant query infrastructure and will offer services that can be hosted on web pages or in Jupyter notebooks. The effort also informs the design of a new, modern landing page that meets the expectations of accessibility and ease of use.

The new query infrastructure is based on nexsciTAP, a component-based, DBMS-agnostic Python implementation of the IVOA Table Access Protocol, developed at NExScI and integrated into the NASA Exoplanet Archive and the NEID Archive, and into the PyKOA Python client. This infrastructure incorporates R-tree spatial indexing, built as memory-mapped files as part of Montage, a software toolkit used to create composite astronomical images. Although R-trees are used most often in geospatial analysis, here they enable searches of the entire KOA archive, an eclectic collection of 100 million records of imaging and spectroscopic data, in 2 seconds, and they speed up spatial searches by x20. The front end is built on the open-source Plotly-Dash framework, which allows users to build an interactive user interface based on a single Python file.
\end{abstract}

\section{Introduction}
The Keck Observatory Archive (KOA)\footnote{\url{https://koa.ipac.caltech.edu}} curates all observations that have been acquired at the W. M. Keck Observatory since operations began in 1994. The holdings include data from all 13 observatory instruments and new data are added in near real-time as they are acquired. Moreover, data volumes and data complexity are increasing rapidly. KOA has begun a program of modernizing the archive infrastructure, now 20 years old and written largely in C. This modernization is driven by two goals:
\begin{enumerate}
    \item Integrate KOA more closely into observatory operations and make data available immediately.
    \item Modernize the software architecture: exploit modern software architecture; 
    support fast queries; enable VO compliance, which is now expected of modern archives  \citep{2017ASPC..512...65A, 2012epsc.conf..626A}.
\end{enumerate}

Item 1 has largely been completed. All data are now archived in near real-time, often less than a minute after acquisition \citep{2022arXiv221202576B}, and a web interface allows observers and collaborators to monitor data during observations \citep{2024SPIE13098E..0JC, 2024arXiv240204528O}. These developments were performed as part of a broader effort to streamline observatory operations through what is called the Data Services Initiative  \citep{2022SPIE12186E..0HB}.
       
Work on item 2 is now underway. This paper reports on prototyping to investigate open-source components needed to develop a new and faster query infrastructure, and when integrated, enable the development of new, scalable services that are not possible with the current architecture. 

\section{VO-Compliant Query Infrastructure}
The nexsciTAP server 
\footnote{\url{https://github.com/Caltech-IPAC/nexsciTAP}} was developed in Python to comply with the International Virtual Observatory Virtual Observatory Table Access Protocol (TAP) \footnote{\url{https://www.ivoa.net/documents/TAP/}} and to ensure data access rights. It has been implemented in the NEID Archive \footnote{\url{https://neid.ipac.caltech.edu/search.php}}, the NASA Exoplanet Archive \footnote{\url{https://exoplanetarchive.ipac.caltech.edu/}}, and in the KOA Python client, PyKOA \footnote{\url{https://koa.ipac.caltech.edu/UserGuide/PyKOA/TAPClients.html}}. Now, it will provide the core query infrastructure for KOA. Figure 1 summarizes its design. The architecture is database-agnostic and enables translation between Astronomy Data Query Language (ADQL), a dialect of SQL to support astronomy \footnote{\url{https://www.ivoa.net/documents/latest/ADQL.html}}, and native SQL itself.

\articlefigure[width=0.65\textwidth]{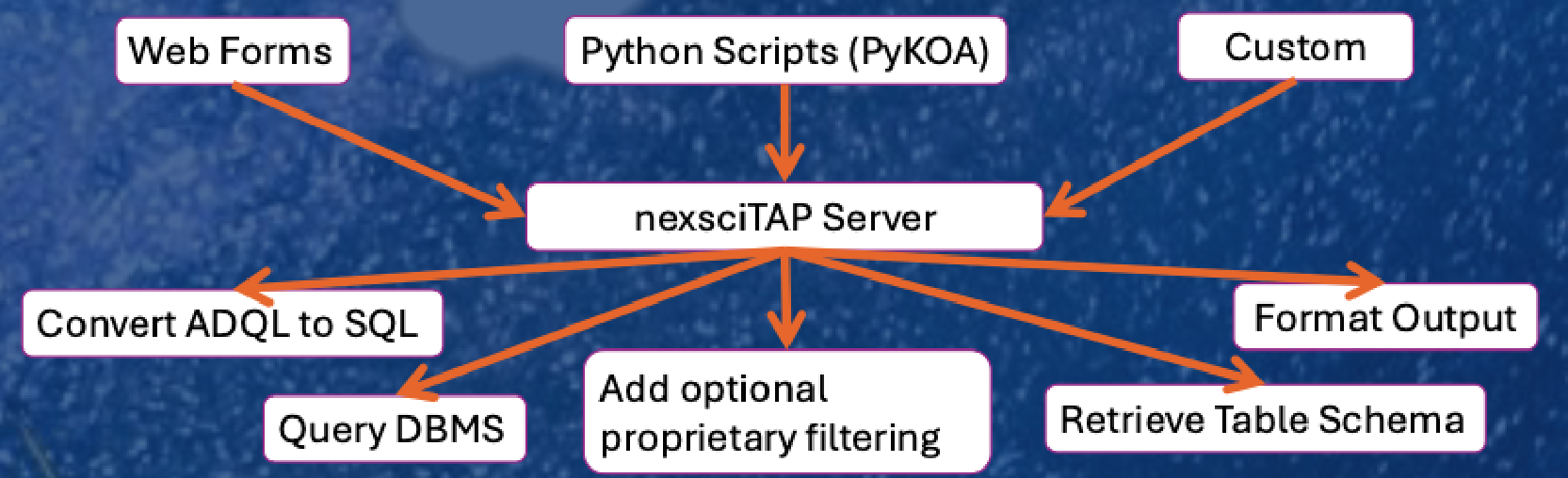}{P412_f1}{The Design of nexsciTAP } 

\section{R-tree Indexing Scheme}
R-trees are tree data structures used for indexing multidimensional information, such as geographical coordinates, rectangles, or polygons. Briefly, they involve grouping nearby objects and representing them with their "minimum bounding rectangle" in the next higher level of the tree \citep{10.1145/602259.602266}. R-trees indices have been implemented in the Montage Image Mosaic Engine as memory-mapped files \footnote{\url{https://github.com/Caltech-IPAC/Montage}}. When applied to KOA data, they search KOA's 6 million files in 2 seconds, a factor of 20 faster than in the current architecture. Performance will scale as the archive grows and can dynamically update as data are added.

\section{Plotly-Dash Dashboard Framework}
The Python-based Plotly-Dash framework\footnote{\url{https://dash.plotly.com/}} is in active use by large commercial businesses, has a large community supporting it, and  accepts contributions from developers. Its architecture offers many benefits in the development of interfaces. In Dash, callbacks handle all events driven by user interaction. The design frees developers to focus on their applications without generating unwieldy code bases. The code itself contains no direct reference to JavaScript or communication with the browser. A major advantage of this approach is in its flexibility: the "back end" can equally be Python run on the user's desktop from the command-line, a Jupyter Notebook page, or a WSGI web service running under Apache.  

\section{Image Visualization}
Plotly-Dash does not contain a useful astronomy image visualizer. We therefore developed one by wrapping a React component around the {\tt mViewer} visualization module in Montage Image Mosaic Engine \citep{2017PASP..129e8006B}. It provides an adaptive histogram-based image display optimized for astronomy, scales symbols by brightness, and repaints coordinate grids on zoom. 

\section{Data Discovery Service}
The KOA Data Discovery Service is a demonstration of the type of service enabled with the new architecture. It integrates nexsciTAP, R-tree indexing, Plotly-Dash, and the Montage visualizer to return inventories of the entire archive in seconds. Figure 2 shows the architecture of this service, which will be publicly released for beta testing in January 2025. 

\section{Conclusion}
The success of the efforts described in this paper has shown that the architectural approach will support KOA in the coming years. We plan to deploy it operationally in late 2025, along with a new KOA landing page. 

\articlefigure[width=0.65\textwidth]{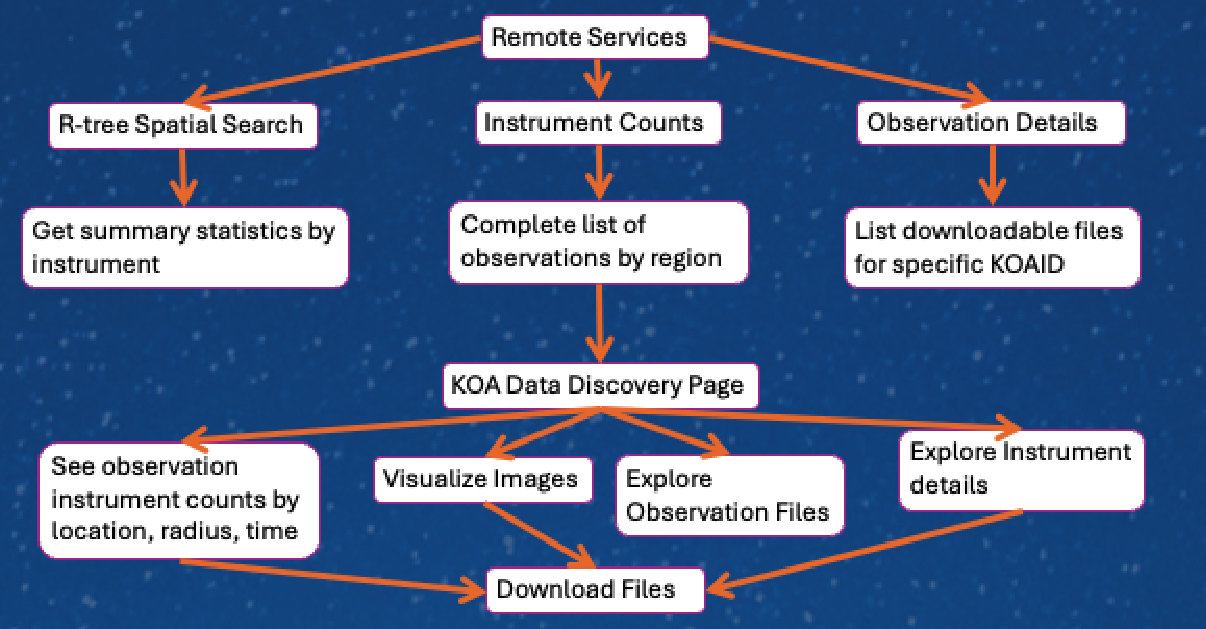}{P412_f2}{Design of the Data Discovery Service}

\acknowledgements The Keck Observatory Archive (KOA) is a collaboration between
the NASA Exoplanet Science Institute (NExScI) and the W. M. Keck Observatory (WMKO). NExScI is sponsored by NASA’s Exoplanet Exploration Program and operated by the California Institute of Technology in coordination with the Jet Propulsion Laboratory (JPL).\

The observatory was made possible by the generous financial support of the W. M. Keck Foundation. The authors wish to recognize and acknowledge the very significant cultural role and reverence that the summit of Mauna Kea has always had within the indigenous Hawaiian community. We are most fortunate to have the opportunity to conduct observations from this mountain.

\bibliography{P412}

\end{document}